\documentclass[reprint, prl, superscriptaddress, longbibliography]{revtex4-1}

\usepackage{amsthm, amssymb, amsmath, hyperref}
\usepackage{CJK}

\newtheorem{theorem}{Theorem}

\theoremstyle{named}
\newtheorem*{dl}{Detectability Lemma}

\DeclareMathOperator{\arccosh}{arccosh}
\DeclareMathOperator{\poly}{poly}

\begin{document}

\begin{CJK*}{UTF8}{}

\title{Correlation Length versus Gap in Frustration-Free Systems}

\CJKfamily{gbsn}

\author{David Gosset}

\email{dngosset@gmail.com}

\affiliation{Walter Burke Institute for Theoretical Physics, California Institute of Technology, Pasadena, California 91125, USA}
\affiliation{Institute for Quantum Information and Matter, California Institute of Technology, Pasadena, California 91125, USA}

\author{Yichen Huang (黄溢辰)}

\email{ychuang@caltech.edu}

\affiliation{Institute for Quantum Information and Matter, California Institute of Technology, Pasadena, California 91125, USA}

\date{\today}

\begin{abstract}

Hastings established exponential decay of correlations for ground states of gapped quantum many-body systems. A ground state of a (geometrically) local Hamiltonian with spectral gap $\epsilon$ has correlation length $\xi$ upper bounded as $\xi=O(1/\epsilon)$. In general this bound cannot be improved. Here we study the scaling of the correlation length as a function of the spectral gap in \textit{frustration-free} local Hamiltonians, and we prove a tight bound $\xi=O(1/\sqrt\epsilon)$ in this setting. This highlights a fundamental difference between frustration-free and frustrated systems near criticality. The result is obtained using an improved version of the combinatorial proof of correlation decay due to Aharonov, Arad, Vazirani, and Landau.

\end{abstract}

\maketitle

\end{CJK*}

Exponential decay of correlations is a basic feature of the ground space in gapped quantum many-body systems. The setting is as follows. We consider a geometrically local Hamiltonian $H$ which acts on particles of constant dimension $s$; i.e., the Hilbert space is $\left(\mathbb{C}^s\right)^{\otimes n}$, where $n$ is the total number of particles. The particles are located at the sites of a finite lattice (of some arbitrary dimension). We write the Hamiltonian as
\[
H=\sum_{i} H_{i},
\]
where distinct terms $H_i,H_j$ are supported on distinct subsets of particles. Here the support of a term $H_i$ is the set of particles on which it acts nontrivially. We assume that $H$ has constant range $r$ with respect to the usual distance function $d$ on the lattice, the shortest path metric. This means that the diameter of the support of each term $H_i$ is upper bounded by $r$ (e.g., $r=2$ for nearest-neighbor interactions). Without loss of generality we assume that the smallest eigenvalue of each term $H_i$ is equal to zero, and that $\|H_i\|\leq 1$.

If $H$ has a unique ground state $|\psi\rangle$ and spectral gap $\epsilon$, connected correlation functions decay exponentially as a function of distance \cite{Has04a, NS06, HK06}. In particular,
\begin{equation} \label{decay}
|\langle\psi|AB|\psi\rangle-\langle\psi|A|\psi\rangle\langle\psi|B|\psi\rangle|\le C \|A\|\|B\|e^{-d(A,B)/\xi}
\end{equation}
holds with
\begin{equation} \label{bound}
\xi=O(1/\epsilon),
\end{equation}
where $d(A,B)$ denotes the distance between the supports of two (arbitrary) local observables $A,B$, and $C$ is a positive constant which depends on $r$ and the lattice. In the transverse field Ising chain the scaling $\xi=\Theta(1/\epsilon)$ is achieved \cite{P70, Con94}, which shows that the upper bound on $\xi$ in Eq. \eqref{bound} cannot be improved.

In gapped systems with (exactly) degenerate ground states, a modification of Eq. (\ref{decay}),
\begin{equation} \label{deg}
|\langle\psi|AB|\psi\rangle-\langle\psi|AGB|\psi\rangle|\le C\|A\|\|B\|e^{-d(A,B)/\xi},
\end{equation}
holds with $\xi=O(1/\epsilon)$ for any ground state $|\psi\rangle$ \cite{Has04b}, where $G$ is the projector onto the ground space. An overview of these results and the proof techniques used to obtain them is given in Ref. \cite{Local10}.

Here we specialize to \textit{frustration-free} geometrically local Hamiltonians.  Frustration-freeness means that any ground state of $H$ is also in the ground space of each term $H_i$. Since we assume that $H_i$ has smallest eigenvalue zero, this means that any ground state $|\psi\rangle$ of $H$ satisfies $H_i|\psi\rangle=0$ for all $i$. The ground energy of $H$ is therefore zero and in general the ground space may be degenerate. The spectral gap $\epsilon$ of $H$ is defined to be its smallest nonzero eigenvalue. Henceforth we assume (without loss of generality \footnote{For a frustration-free local Hamiltonian $H=\sum_i H_i$ we may consider a related Hamiltonian $H'=\sum_i \Pi_i$, where $\Pi_i$ projects onto the range of $H_i$. The zero energy ground spaces of $H$ and $H'$ are the same, and their spectral gaps $\epsilon,\epsilon'$ are related as follows. Letting $a>0$ be such that $a\Pi_i \leq H_i$ for all $i$, and noting that $H_i\leq \Pi_i$, we obtain $a\epsilon'\leq \epsilon\leq \epsilon'$.}) that each term $H_i$ in the Hamiltonian is a projector, i.e., $H_i^2=H_i$. 

Frustration-free Hamiltonians are widely studied in physics and quantum computation.

In quantum complexity theory there is a powerful analogy between local Hamiltonians and constraint satisfaction problems \cite{Kitaev02}. If we view the terms $H_i$ as constraints, then computing the ground energy is a quantum constraint satisfaction problem. In this language frustration-freeness means that all constraints can be satisfied simultaneously. (Classical) satisfiability is a computational problem in which one is given a set of constraints and asked to determine if they are simultaneously satisfiable. Its quantum analogue, defined by Bravyi \cite{Bravyi06} (see also Refs. \cite{GN13,Laumann10,LaumannProd, BravyiMoore, quantumLovasz, Shearer}), is a computational problem in which one is given a local Hamiltonian and asked to determine if it is frustration-free. 

The study of quantum states where the entanglement structure is encoded in a tensor network has led to a paradigm shift in quantum many-body physics (see, e.g., Ref. \cite{WenBook}). There is a close connection between such tensor network states and frustration-free Hamiltonians: for matrix product states or projected entangled pair states one can construct a frustration-free parent Hamiltonian as a sum of local projectors, where each projector annihilates the local reduced density matrix of the state \cite{FNW92,PVWC07,peps}.  

Gapped frustration-free systems include widely studied spin chains such as the AKLT model \cite{AKLT87} and the spin-$1/2$ ferromagnetic $XXZ$ chain (with kink boundary conditions) \cite{GW95, ASW95, KN97}. A prevalent strategy to study topological phases in two and higher spatial dimensions is to construct exactly solvable models, such as the toric code \cite{Kit03} and, more generally, the quantum double \cite{Kit03} and string net \cite{LW05} models. Almost all such models are gapped and frustration-free (commuting, even). Gapless frustration-free systems include the spin-$1/2$ ferromagnetic Heisenberg chain and the Rokhsar-Kivelson quantum dimer model \cite{RS88}.  Two recent papers construct gapless frustration-free spin chains in which the half-chain entanglement entropy diverges in the thermodynamic limit $n\rightarrow\infty$: a spin-$1$ example based on parenthesized expressions (with $\log(n)$ divergence) \cite{spin1} and a higher spin generalization ($\sqrt{n}$ divergence) \cite{MS14}.  The classification of spin-$1/2$ chains in Ref. \cite{BG15} provides many further examples of gapped and gapless frustration-free systems.

We establish the following tight upper bound on correlation length in frustration-free systems near criticality (i.e., in the limit $\epsilon\rightarrow 0$).

\begin{theorem}
Suppose $H$ is a frustration-free geometrically local Hamiltonian with spectral gap $\epsilon$. Then decay of correlations \eqref{deg} holds with
\begin{equation}
\xi=O(1/\sqrt\epsilon)
\end{equation}
for any ground state $|\psi\rangle$ of $H$.
\label{thm:main}
\end{theorem}
In this theorem $C$ in the bound \eqref{deg} is an absolute constant, while the constant hidden in the big-O notation depends only on the interaction range $r$ and a parameter $g$ defined by
\begin{equation}
g=\max_i|\{H_j:[H_i,H_j]\neq0\}|.
\label{eq:gdef}
\end{equation}
($g$ itself only depends on $r$ and the geometry of the lattice.)

Theorem \ref{thm:main} may be of interest for at least three reasons. First, for matrix product states or projected entangled pair states, it implies an upper bound on the energy gap of the parent Hamiltonian in terms of the correlation length of the state. Second, while frustration-free models such as those discussed above seem to be representative of many \text{gapped} phases of matter, our result states that gapless frustration-free systems cannot exhibit critical phenomena with, e.g., $\xi=\Theta(1/\epsilon)$. This may be relevant to understanding possible scaling limits of critical frustration-free systems, an issue which has been raised in Refs. \cite{spin1,MS14,BG15}. The third and final reason, explained below, is its potential relevance to the area law for the entanglement entropy in one-dimensional spin systems.

There is a folklore argument which relates correlation length and entanglement entropy in one-dimensional spin systems. Combined with Theorem \ref{thm:main}, this argument suggests an improved area law for ground-state entanglement in frustration-free gapped one-dimensional systems. The area law states that the ground state entanglement entropy $S$ for a contiguous region of a (possibly frustrated) one-dimensional spin system is upper bounded by a constant that is independent of the size of the region  (but which depends on the energy gap $\epsilon$) \cite{Has07}. In particular \cite{AKLV13, area}\footnote{We use a tilde to hide a polylogarithmic factor, e.g., $\tilde O(x):=O(x\poly\log x)$}
\begin{equation} \label{eq:arealaw}
S=\tilde O(1/\epsilon).
\end{equation}
This result agrees with the non-rigorous folklore argument that particles should be almost uncorrelated if their distance is beyond a constant multiple of the correlation length. One expects $S=O(\xi)$ because only $O(\xi)$ particles in a neighborhood of a cut should contribute non-negligibly to the entanglement across the cut. Thus, (\ref{eq:arealaw}) is suggested by (\ref{bound}) up to a polylogarithmic prefactor. We emphasize that it is at least very challenging (if not impossible \cite{Has15}) to make this argument rigorous. Indeed, the connection between entanglement entropy and correlation length has only been proved in a weaker sense: $S=\exp[\tilde O(\xi)]$ \cite{BH13}. Nevertheless it is interesting to note that the same argument along with Theorem \ref{thm:main} indicates that the stronger bound
\begin{equation}
S\overset{?}{=}O(1/\sqrt\epsilon)
\label{eq:SFF}
\end{equation}
might hold in frustration-free systems. This point, and further evidence for Eq. \eqref{eq:SFF}, is discussed in Ref. \cite{GM15}.

We now turn to the proof of Theorem \ref{thm:main}. Reference \cite{AAVL11} gives a combinatorial proof of correlation decay for frustration-free Hamiltonians. That proof gives an upper bound $\xi=O(1/\epsilon)$. To prove Theorem \ref{thm:main}, we modify the argument from Ref. \cite{AAVL11} using Chebyshev polynomials. In the field of Hamiltonian complexity \cite{Osb12, GHLS14}, Chebyshev polynomials have been used to prove area laws for the entanglement entropy in the ground states of one-dimensional gapped systems \cite{ALV12, AKLV13, area}.  It is thus not surprising that they are useful in the present context \footnote{As in those previous works, below we make use of an operator [denoted $Q_m(P^\dagger P)$] which is an approximate ground space projector (AGSP). However, the AGSP used here is different from those used in Refs. \cite{ALV12, AKLV13, area}. In our proof we do not need an AGSP with small entanglement rank, which gives us greater freedom.}.

\begin{proof}---The first part of the proof follows Ref. \cite{AAVL11}. For completeness, we review the necessary material from that paper; we indicate below where this proof differs. Here we use a slightly different version of the detectability lemma \cite{AAVL11}. A proof of this version is given in Ref. \cite{AAV16}.

\begin{dl} [\cite{AAVL11, AAV16}]
Let $H=\sum_i H_i$ be a frustration-free local Hamiltonian with ground space projector $G$ and spectral gap $\epsilon$. Choose some ordering of the terms $H_i$ and let $P=\prod_i(1-H_i)$, where the product is taken with respect to this ordering. Then,
\begin{equation} \label{DL}
\|P-G\|\le1\big/\sqrt{1+\epsilon/g^2}
\end{equation}
where $g$ is given by Eq. \eqref{eq:gdef}.
\end{dl}

The result (\ref{DL}) holds for any order of the projectors $(1-H_i)$ in the definition of $P$.  Following Ref. \cite{AAVL11}, we fix a particular order as follows. It will be useful to define an ``interaction graph'' with a vertex for each term $H_i$ and an edge between two vertices if the corresponding terms do not commute. Note that $g$ is the maximum degree of this interaction graph. 

We first partition the projectors $(1-H_i)$ into a constant number $c$ of layers (sets) such that any two projectors within a given layer commute. This partition can be obtained from a proper vertex coloring of the interaction graph using $c$ colors (no two vertices with the same color share an edge). Since any graph with maximum degree $\Delta$ has such a coloring with $\Delta+1$ colors we get $c\leq g+1$. For example, for nearest-neighbor interactions in one dimension, each projector $H_i$ has support on particles $i,i+1$, and we may take $c=2$ (with one layer consisting of all projectors with even values of $i$, and the other layer corresponding to odd values of $i$). After fixing the layers, we then choose some (arbitrary) ordering of them; e.g., in one dimension we might take the odd layer to be first and the even layer to be second. Finally, we take $P=L_c\cdots L_2 L_1$, where $L_j$ is the product of all projectors $(1-H_i)$ in layer $j$.
Choosing $P$ in this way we have \cite{AAVL11}
\begin{equation} \label{eat}
\langle\psi|A(P^\dag P)^mB|\psi\rangle=\langle\psi|AB|\psi\rangle
\end{equation}
for $m<d(A,B)/((2c-1)(r-1))$. To see why Eq. \eqref{eat} holds, we view $(P^\dag P)^m$ as consisting of $(2c-1)m$ layers. The reader may find it helpful to look at Fig. 4 in Ref. \cite{AAVL11}. Note that $(1-H_i)|\psi\rangle=|\psi\rangle$; i.e., any projector $(1-H_i)$ acts as the identity on a ground state $|\psi\rangle$. Likewise, for any term $H_i$ with support disjoint from that of $A$ we have $(1-H_i)A|\psi\rangle=|\psi\rangle$. More generally, in the expression $\langle\psi|A(P^\dag P)^m$ we may replace many of the projectors with the identity; the ones that remain are said to be in the causal cone of $A$.  Each layer reduces the distance between $B$ and the causal cone of $A$ by at most $(r-1)$. So if $m(2c-1)(r-1)<d(A,B)$ then every projector in the causal cone of $A$ acts trivially on $B|\psi\rangle$ and Eq. \eqref{eat} follows.

At this point we depart from the proof given in Ref. \cite{AAVL11}, using ideas from Ref. \cite{AKLV13}. Equation \eqref{eat} directly implies that for any degree-$m$ polynomial $Q_m(x)$ with $Q_m(1)=1$ and $m<d(A,B)/((2c-1)(r-1))$ we have
\begin{equation} \label{eat2}
\langle\psi|AQ_m(P^\dag P)B|\psi\rangle=\langle\psi|AB|\psi\rangle.
\end{equation}
We choose $Q_m$ to be a rescaled and shifted Chebyshev polynomial defined by
\begin{equation}
Q_{m}(x)=\frac{T_m\left(\frac{2x}{1-\delta}-1\right)}{T_m\left(\frac{2}{1-\delta}-1\right)}, \quad \text{where}~\delta=\frac{\epsilon}{g^2+\epsilon},
\label{eq:Qm}
\end{equation}
and $T_m(x)=\cos(m\arccos x)=\cosh(m\arccosh x)$ is the standard (degree-$m$) Chebyshev polynomial of the first kind. The function $Q_m$ is a degree-$m$ polynomial with $Q_m(1)=1$ and \cite{AKLV13}
\begin{equation}
|Q_m(x)|\le2e^{-2m\sqrt\delta} \quad \text{for}~0\leq x\leq 1-\delta.
\label{eq:Qmprop}
\end{equation}
Equation \eqref{eq:Qmprop} follows from Eq. \eqref{eq:Qm} and the facts that \cite{AKLV13}
\begin{align}
&|T_m(x)|\le1&&\text{for}~|x|\le1\nonumber\\
&T_m(x)>\frac{1}{2}e^{2m\sqrt{(x-1)/(x+1)}}&&\text{for}~x>1.\nonumber
\end{align}

Since $G$ projects onto the $+1$ eigenspace of the positive semidefinite operator $P^\dagger P$, we have $P^\dagger P-G\geq 0$. Using the detectability lemma we get 
\[
\|P^\dagger P-G\|=\|P-G\|^2\leq \frac{1}{1+\epsilon/g^2}=1-\delta
\]
and, therefore, we have the operator inequality
\begin{equation}
0 \leq P^\dag P-G \le (1-\delta)\cdot 1.
\label{eq:PP}
\end{equation}
Now, let $G^\perp=1-G$. Again, using the fact that $G$ projects onto the $+1$ eigenspace of $P^\dagger P$ and the fact that $Q_m(1)=1$, we have
\begin{eqnarray} \label{clo}
Q_m(P^\dag P)-G&=&G^\perp(Q_m(P^\dag P)-G)G^\perp\nonumber\\
&=&G^\perp Q_m(P^\dag P-G)G^\perp.
\end{eqnarray}

Using Eq. (\ref{eat2}) and then Eq. (\ref{clo}) we have, for all $m<d(A,B)/((2c-1)(r-1))$,
\begin{align}
&|\langle\psi|AB|\psi\rangle-\langle\psi|AGB|\psi\rangle|\nonumber\\
&=|\langle\psi|A(Q_m(P^\dagger P)-G)B|\psi\rangle|\nonumber\\
&\leq \|A\|\|B\|\|Q_m(P^\dag P)-G\|\nonumber\\
& \leq \|A\|\|B\|\|Q_m(P^\dag P-G)\|\nonumber\\
& \le2\|A\|\|B\|\exp\left(-2m\sqrt{\frac{\epsilon}{g^2+\epsilon}}\right),\label{eq:last}
\end{align}
where in the last inequality we used Eqs. \eqref{eq:Qmprop} and \eqref{eq:PP}. We now choose $m$ to be the largest integer less than $d(A,B)/((2c-1)(r-1))$. Substituting the bound $m\geq d(A,B)/((2c-1)(r-1))-1$ in Eq. \eqref{eq:last} and using the fact that $\epsilon/(g^2+\epsilon)\leq 1$, we arrive at the desired bound \eqref{deg} with $C=2e^2$ and
\[
\xi=\frac{(2c-1)(r-1)}{2}\sqrt{\frac{g^2+\epsilon}{\epsilon}}=O(1/\sqrt{\epsilon}).
\]
\end{proof}

A simple example shows that the upper bound  on $\xi$ in Theorem \ref{thm:main} cannot be improved. Consider the spin-$1/2$ ferromagnetic $XXZ$ chain with kink boundary conditions \cite{GW95, ASW95, KN97}. The Hamiltonian for the chain of length $n$ can be written as a sum of projectors
\begin{equation} \label{proj}
H(q)=\sum_{i=1}^{n-1}|\phi(q)\rangle\langle\phi(q)|_{i,i+1},\quad|\phi(q)\rangle=\frac{q|1 0\rangle-|01\rangle}{\sqrt{q^2+1}}
\end{equation}
where $0< q<1$ and $|0\rangle,|1\rangle$ are spin up and down, respectively. The spectral gap of $H(q)$ is given by \cite{KN97}
$$\epsilon=1-\frac{2}{q+q^{-1}}\cos(\pi/n)$$
and vanishes as $q\rightarrow 1$ and $n\rightarrow\infty$. 

The total magnetization $M=\sum_{i=1}^n\frac{1}{2}(1-\sigma_i^z)$ is conserved. It can be verified by a direct computation that 
\[
|\psi_1\rangle=\left(\frac{1-q^{2}}{1-q^{2n}}\right)^{1/2}\sum_{j=1}^n q^{j-1} \sigma_j^{x}|0 0 \ldots 0\rangle
\]
satisfies $H(q)|\psi_1\rangle=0$ and is the unique ground state in the symmetry sector where $M$ has eigenvalue $1$. Let $A=\frac{1}{2}(1-\sigma_1^z)$ and $B=\frac{1}{2}(1-\sigma_j^z)$ for some $j>1$, so that $d(A,B)=j-1$. Then $\langle\psi_{1}|AB|\psi_{1}\rangle=0$ and so
\begin{eqnarray}
&&|\langle\psi_{1}|AB|\psi_{1}\rangle-\langle\psi_{1}|AGB|\psi_{1}\rangle|\nonumber\\
&&=|\langle\psi_{1}|A|\psi_1\rangle\langle \psi_1|B|\psi_{1}\rangle|\nonumber\\
&&=\left(\frac{1-q^2}{1-q^{2n}}\right)^2 q^{2d(A,B)},
\label{eq:xxzcor}
\end{eqnarray}
where, in the first equality, we used the fact that $A$ and $B$ commute with $M$. For simplicity we now take the limit $n\rightarrow\infty$. If we suppose Eq. \eqref{deg} holds for some $C$ and $\xi$, then Eq. \eqref{eq:xxzcor} implies
\[
(1-q^2)^2 q^{2d(A,B)}\leq Ce^{-d(A,B)/\xi}.
\]
Taking logs on both sides and using the fact that $\xi$ does not depend on $d(A,B)$ gives the desired lower bound
\[
\xi\geq \frac{1}{-2\ln{q}}=\frac{1}{-2\ln(1-O(\sqrt{\epsilon}))}=\Omega(1/\sqrt{\epsilon}),
\]
where in the second step we used the fact that $\epsilon=1-2/(q+q^{-1})$ (in the limit $n\rightarrow\infty$).

\emph{Remark.---}The $XXZ$ chain also seems to nicely illustrate the optimality of the bound \eqref{DL} in the detectability lemma. The bound states that $1-\|P-G\|=\Omega(\epsilon)$, while we found using numerical diagonalization that for the $XXZ$ chain \eqref{proj} $1-\|P-G\|$ is exactly equal to $\epsilon$, for all choices of $q$ and $n$ that we tried (we used the aforementioned two-layer ordering of projectors in the definition of $P$). Presumably this equality holds for all $0<q\leq1$ and $n\geq 2$.

\begin{acknowledgments}
We thank Spiros Michalakis and John Preskill for interesting discussions. We acknowledge funding provided by the Institute for Quantum Information and Matter, an NSF Physics Frontiers Center (NSF Grant No. PHY-1125565) with support of the Gordon and Betty Moore Foundation (GBMF-12500028).

D. G. and Y. H. contributed equally to this work; the author ordering is alphabetical.
\end{acknowledgments}

\bibliography{correlation}

\end{document}